\documentclass{aa}
\usepackage{psfig}
\usepackage{times}
\def\mag{\hbox{$^{\rm m}$}}
\begin{document}

\thesaurus{06               
           (02.19.1;        
            08.03.4;        
            08.06.2;        
            09.08.1;        
            09.10.1;        
            13.09.6)}       

\title{ISO Spectroscopy of the Young Bipolar Nebulae S106 IR and Cep~A East
\thanks{Based on observations with ISO, an ESA project with instruments
        funded by ESA Member States (especially the PI countries: France,
        Germany, the Netherlands and the United Kingdom) and with the
        participation of ISAS and NASA.}}

\author{M.E. van den Ancker\inst{1,2} \and A.G.G.M. Tielens\inst{3,4,5} 
        \and P.R. Wesselius\inst{4}}
\institute{
Astronomical Institute ``Anton Pannekoek'', University of Amsterdam, 
 Kruislaan 403, NL--1098 SJ  Amsterdam, The Netherlands \and
Harvard-Smithsonian Center for Astrophysics, 60 Garden Street, MS 42, 
 Cambridge, MA 02138, USA \and 
Kapteyn Astronomical Institute, Groningen University, P.O. Box 800, 
 NL--9700 AV  Groningen, The Netherlands \and
SRON, P.O. Box 800, NL--9700 AV  Groningen, The Netherlands \and
NASA Ames Research Center, MS 245-3, Moffett Field, CA 94035, USA} 

\offprints{M.E. van den Ancker (mario@astro.uva.nl)}
\date{Received <date>; accepted <date>} 

\maketitle

\begin{abstract}
We present the results of ISO SWS and LWS grating scans towards 
the embedded Young Stellar Objects (YSOs) S106 IR and 
Cep~A East. Emission from the pure rotational lines of H$_2$ and  
the infrared fine structure lines of [C\,{\sc ii}], [O\,{\sc i}], 
[S\,{\sc i}], [Si\,{\sc ii}] and [Fe\,{\sc ii}], as well as 
absorption bands due to H$_2$O, CO and CO$_2$ ice were detected toward 
Cep~A. In S106 we detected emission lines of H$_2$, CO, H\,{\sc i}, 
and a large number of ionized species including Fe, O, N, C, Si, S, 
Ne and Ar. S106 also shows many of the infrared PAH bands in emission. 
Excitation temperatures and molecular hydrogen masses were derived 
from the low-lying pure rotational levels of H$_2$ and are 500 and 
730~K and 8 and 3 $\times$ 10$^{-3}$~M$_\odot$ for S106 and Cep~A, 
respectively. Since both objects are expected to have several solar 
masses of H$_2$ in their environment, we conclude that 
in both cases the bulk of the H$_2$ is cooler than a few hundred 
Kelvins. Excitation temperatures and line ratios were compared with 
those predicted by theoretical models for PDRs and dissociative and 
non-dissociative shocks. 
The [S\,{\sc i}] 25.2~$\mu$m/[Si\,{\sc ii}] 34.8~$\mu$m ratio is a 
particularly useful shock versus PDR discriminant and we conclude that 
S106 IR is dominated by PDR emission while Cep~A East has a large shock 
component. From an analysis of the ionic lines in S106 we conclude that 
the central star must have a temperature around 37,000~K, corresponding 
to a spectral type of O8. From its luminosity it is concluded that 
the driving source of Cep~A must also be a massive early-type star. 
The absence of strong high-ionization ionic lines in its ISO 
spectrum shows that Cep~A has not yet created a significant 
H\,{\sc ii} region and must be younger than S106, illustrating the 
process of the clearing of the surroundings of a massive young star.
\end{abstract}

\section{Introduction}
The infrared emission-line spectrum of a Young Stellar Object (YSO) 
is dominated by the interaction of the central star with the 
remnants of the clouds from which it formed. The intense UV radiation 
generated by accretion as well as by the central star itself causes 
dissociation of molecular material close to the YSO and ionizes 
much of the atomic material, giving rise to typical nebular and 
recombination lines. Furthermore, the interaction of the UV field 
with the remnants of the star's natal cloud will produce a
photodissociation region (PDR; see Hollenbach \& Tielens 1999
for a comprehensive review). In a PDR, heating of the gas occurs 
by collisions with electrons, photoelectrically ejected from grain 
surfaces. The gas in the surface regions of PDRs reaches 
temperatures of typically 500~K (e.g. Tielens \& Hollenbach 1985), 
producing a distinctive infrared spectrum due to collisionally 
excited low-lying levels of many molecular and neutral species.

Another phenomenon associated with YSOs are strong stellar winds, 
often collimated into a bipolar outflow. These will cause a shock 
as they drive a supersonic wave into the surrounding molecular 
cloud, heating it in the process. Shocks are usually divided 
into two distinct categories: J- or Jump-shocks, and C- or 
Continuous-shocks. In J-shocks viscous heating of the neutrals occurs in 
a thin shock front in which radiative cooling is insignificant, while the 
post-shock gas is heated to several times 10$^4$ degrees, dissociating 
all molecular material (e.g. Hollenbach \& McKee 1989). Cooling of the 
post-shock gas occurs through atomic fine-structure lines as well as 
through re-formation of molecules. 
In contrast, C-shocks are magnetized, non-dissociative shocks in which 
the physical conditions change more gradually from their pre- to post-shock 
values and cooling is mainly through radiation from molecular material 
(e.g. Kaufman \& Neufeld 1996). 
If the temperatures in a C-shock become sufficiently high to start to 
dissociate molecules, the cooling through the molecular lines diminishes, 
and the shock temperature increases until it turns into a J-shock. 
Shocks with a shock velocity larger than 40~km~s$^{-1}$ are usually 
J-shocks, whereas slower shocks are usually of C-type (Chernoff et al. 1982).
In this paper we will illustrate the power of infrared spectroscopy 
to study the above mentioned phenomena using two young bipolar 
nebulae, S106 and Cep~A.
\begin{figure}[t]
\centerline{\psfig{figure=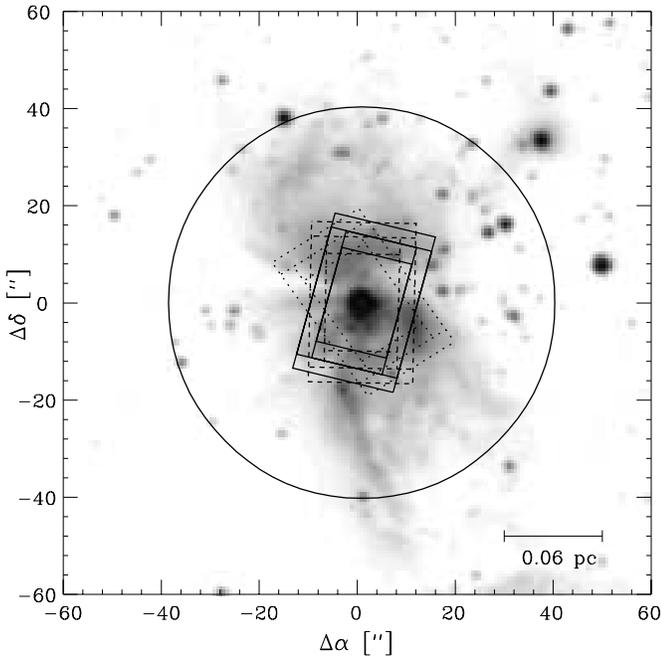,width=8.7cm,angle=0}}
\caption[]{SWS AOT S01 (solid rectangles), AOT S02 (dashed rectangles), 
AOT S06 (dotted rectangles) and LWS (solid circle; beam average FWHM) 
aperture positions for our measurements of S106 
superimposed on a K'-band image of the region. The rectangles 
indicate the apertures (in increasing size) for SWS bands 1A--2C 
(2.4--12.0~$\mu$m), 3A--3D (12.0--27.5~$\mu$m), 3E (27.5--29.5~$\mu$m) 
and 4 (29.5--40.5~$\mu$m).}
\end{figure}
\begin{figure}[t]
\centerline{\psfig{figure=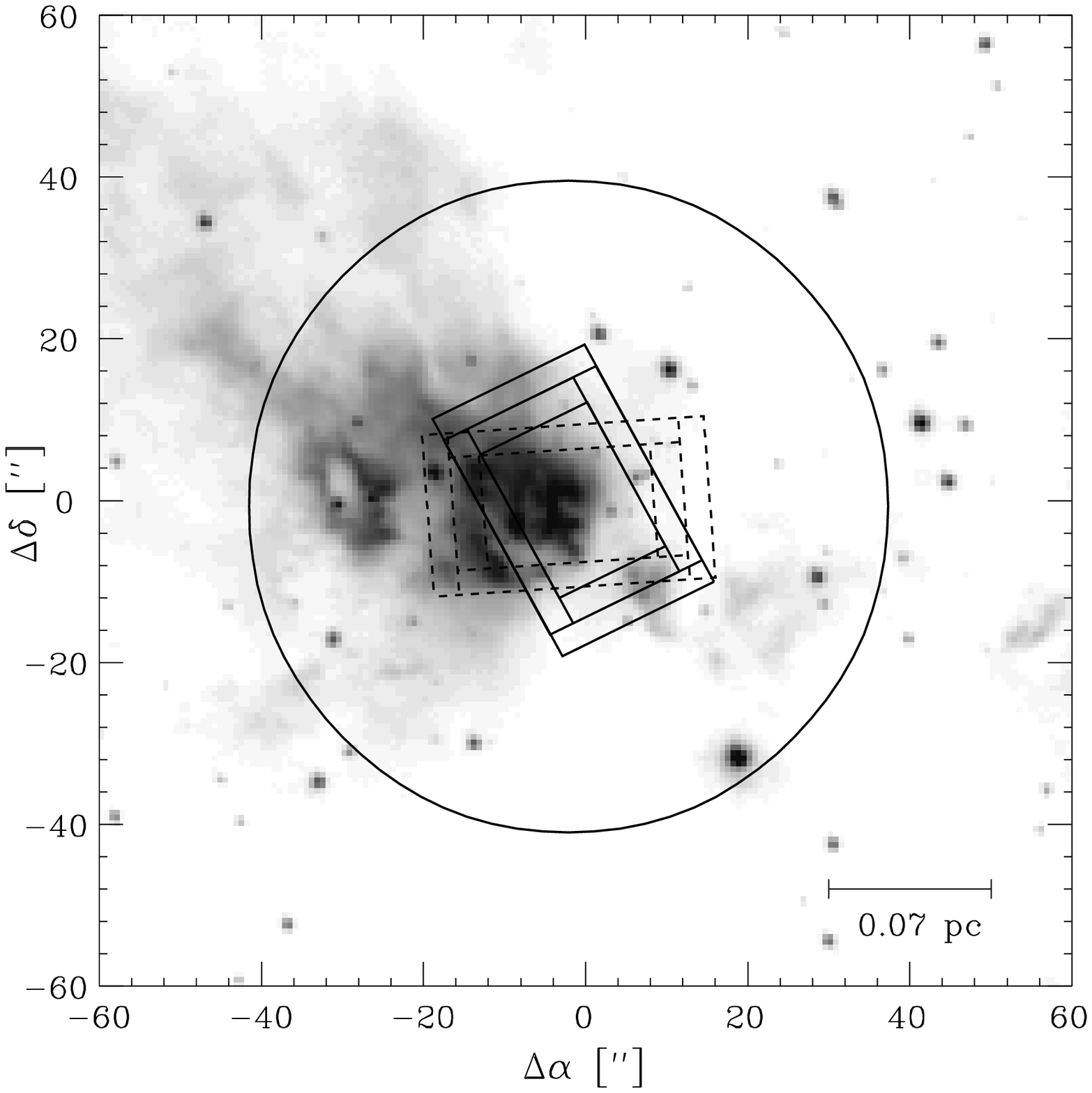,width=8.7cm,angle=0}}
\caption[]{Same as Fig.~2 for Cep~A East.}
\end{figure}

The bipolar nebula S106 is one of the most studied H\,{\sc ii} regions 
in our galaxy. The exciting source of the region is a very young massive 
stellar object, known as either IRS4 in the terminology of Gehrz et al. 
(1982) or IRS3 following Pipher et al. (1976). Around this central source, 
Hodapp \& Rayner (1991) found a cluster of about 160 stars, embedded in 
the molecular cloud surrounding S106. Observed radio 
emission and H\,{\sc i} recombination lines from the region have 
been suggested to arise in a strong ($\dot{M}$ $\approx$ 
2 $\times$ 10$^{-5}$~M$_\odot$~yr$^{-1}$), fast ($\approx$ 200~km~s$^{-1}$), 
stellar wind, driving a shock into the surrounding extended 
molecular cloud (Hippelein \& M\"unch 1981). A dark lane, largely devoid 
of any optical or infrared emission, separates the two lobes of the 
H\,{\sc ii} region. This dark lane has been quoted many times as a
prime example of a large (30\arcsec) circumstellar disk, consisting of 
cool gas and dust, seen nearly edge-on (Bieging 1984; Mezger et al. 1987). 
However, Loushin et al. (1990) showed this structure to be an expanding ring 
of molecular material rather than a protoplanetary disk. Near-infrared 
ro-vibrational emission of H$_2$ is seen to arise just outside the 
H\,{\sc ii} region, suggesting an origin in a PDR (Hayashi et al. 1990).
\begin{figure*}
\centerline{\psfig{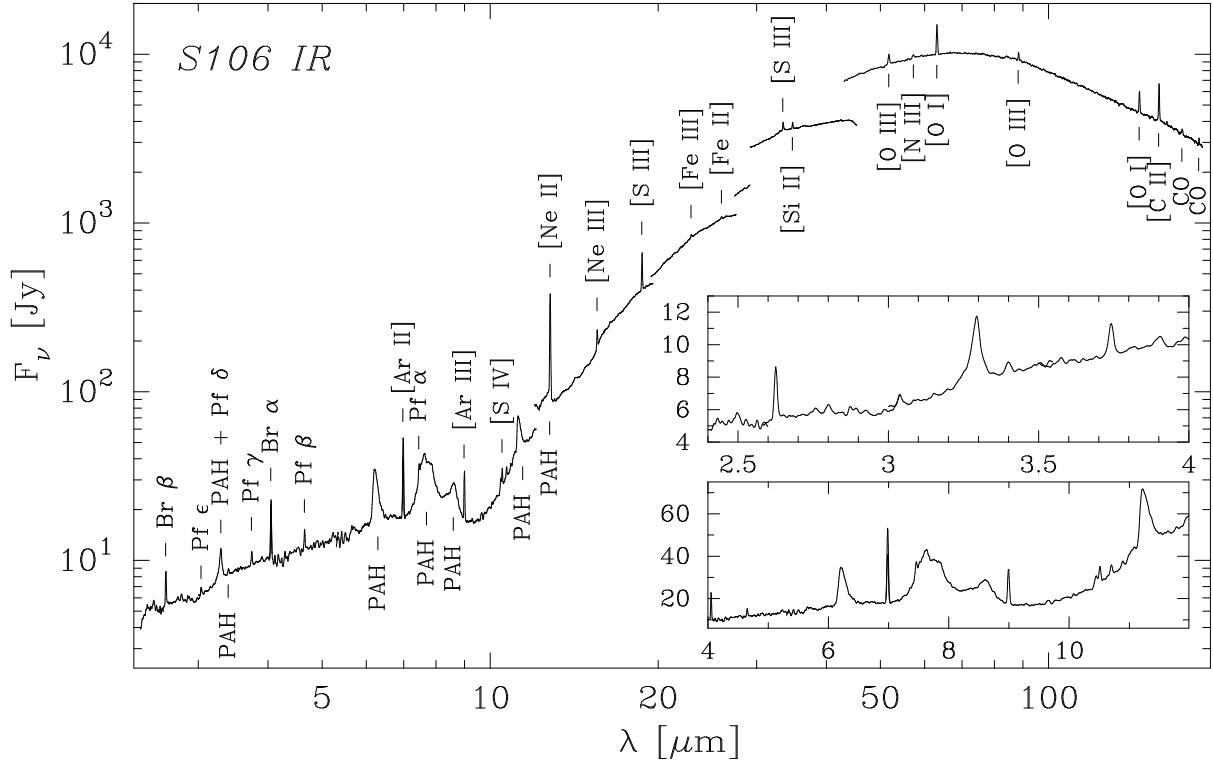}}
\caption[]{Combined ISO SWS/LWS full grating spectrum for S106 IR with 
the most prominent features identified. The insets show enlargements 
of the SWS spectrum in the 2.4--4.0 and 4.0--12.0 micron range.}
\end{figure*}
\begin{figure*}
\centerline{\psfig{figure=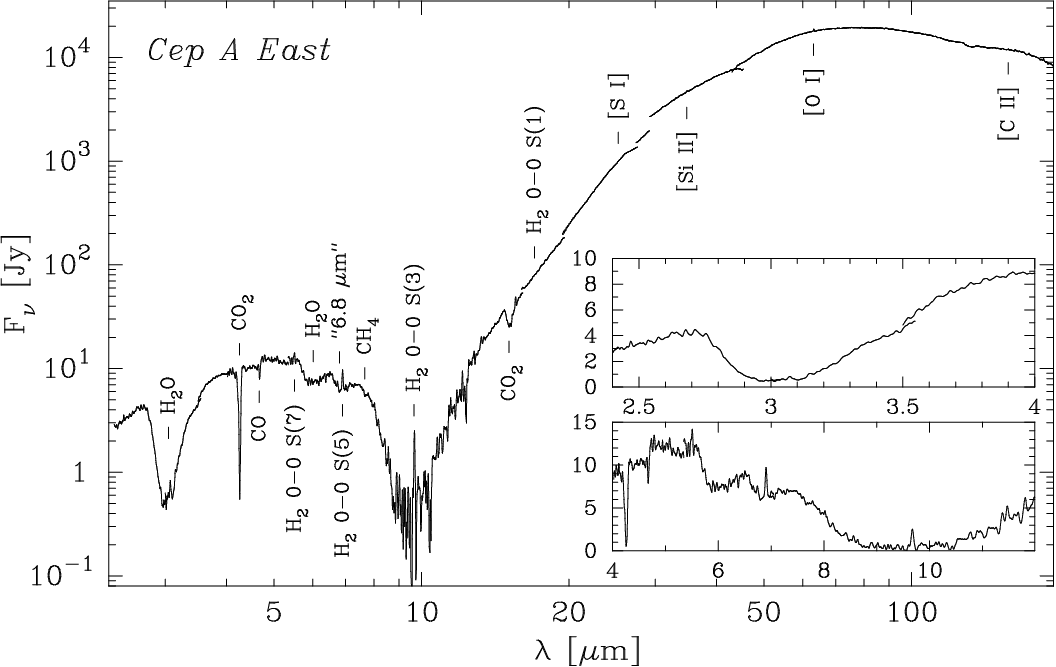,width=16cm,angle=270}}
\caption[]{Same as Fig.~3 for Cep~A East.}
\end{figure*}

Cep~A is another well-known site of recent star formation. It contains a 
luminous (2.5 $\times$ 10$^4$~L$_\odot$; Evans et al. 1981) far-infrared 
source, as well as several fainter infrared sources. The core of the 
Cep~A region remains obscured at optical and infrared wavelengths due to 
massive amounts of extinction, and is the source of an energetic, complex 
molecular outflow (Bally \& Lane 1982). Radio-observations show that it 
contains two strings of about 13 ultracompact H\,{\sc ii} regions, arranged 
in a Y-shape (Hughes 1988). The source IRS 6a (Lenzen et al. 
1984), located in the infrared nebula to the east of the core, is the 
dominant source of the eastern lobe at 20~$\mu$m (Ellis et al. 1990), but 
polarization measurements show that the nebula is illuminated by 
the compact radio source HW-2 (Hughes \& Wouterloot 1984), located 
$\sim$~5\arcsec~ south of IRS 6a (Casement \& McLean 1996). The fact that it 
is not visible at 20~$\mu$m implies an extremely large extinction towards 
HW-2. H$_2$ emission occurs in both the eastern and western parts 
of the nebula (Doyon \& Nadeau 1988). Hartigan et al. (1996) showed that 
the molecular emission to the east appears as a regular jet, whereas that to
the west concentrates in shells, which they proposed to arise in wakes 
from bow shocks surrounding Herbig-Haro objects in the region. 
ISO observations of the western part of the region were 
discussed by Wright et al. (1996), who modelled the observed infrared fine 
structure lines as arising in a planar J-shock with shock velocity 
70--80~km~s$^{-1}$. The same authors modelled the observed molecular hydrogen 
emission as arising in a combination of several C-shocks. This multitude of 
shocks in the region agrees with the explanation of Narayanan \& Walker 
(1996), who reported the presence of multiple episodes of outflow activity 
from the region. Recently, Goetz et al. (1998) presented strong evidence 
for this scenario through near-infrared H$_2$ and [Fe\,{\sc ii}] images of 
Cep~A East showing several distinct regions containing shocks.

In this paper we will present new {\it Infrared Space Observatory} (ISO; 
Kessler et al. 1996) data on infrared fine-structure and molecular emission 
lines from the central region of S106, centered on IRS4, and the eastern 
part of Cep~A, centered on IRS 6a. These data give for the first time 
access to a large spectral range that was unaccessible from ground-based 
observations and contain lines which are decisive probes of the 
excitation conditions. We will show that the observed emission-line 
spectrum can be well explained as arising in the combination of a PDR 
and a H\,{\sc ii} region in the case of S106, whereas this emission is 
shock-excited in the case of Cep~A.

\section{Observations and data}
%
%
\begin{figure*}
\vspace*{-5.5cm}
\hspace*{-0.5cm}
\centerline{\psfig{figure=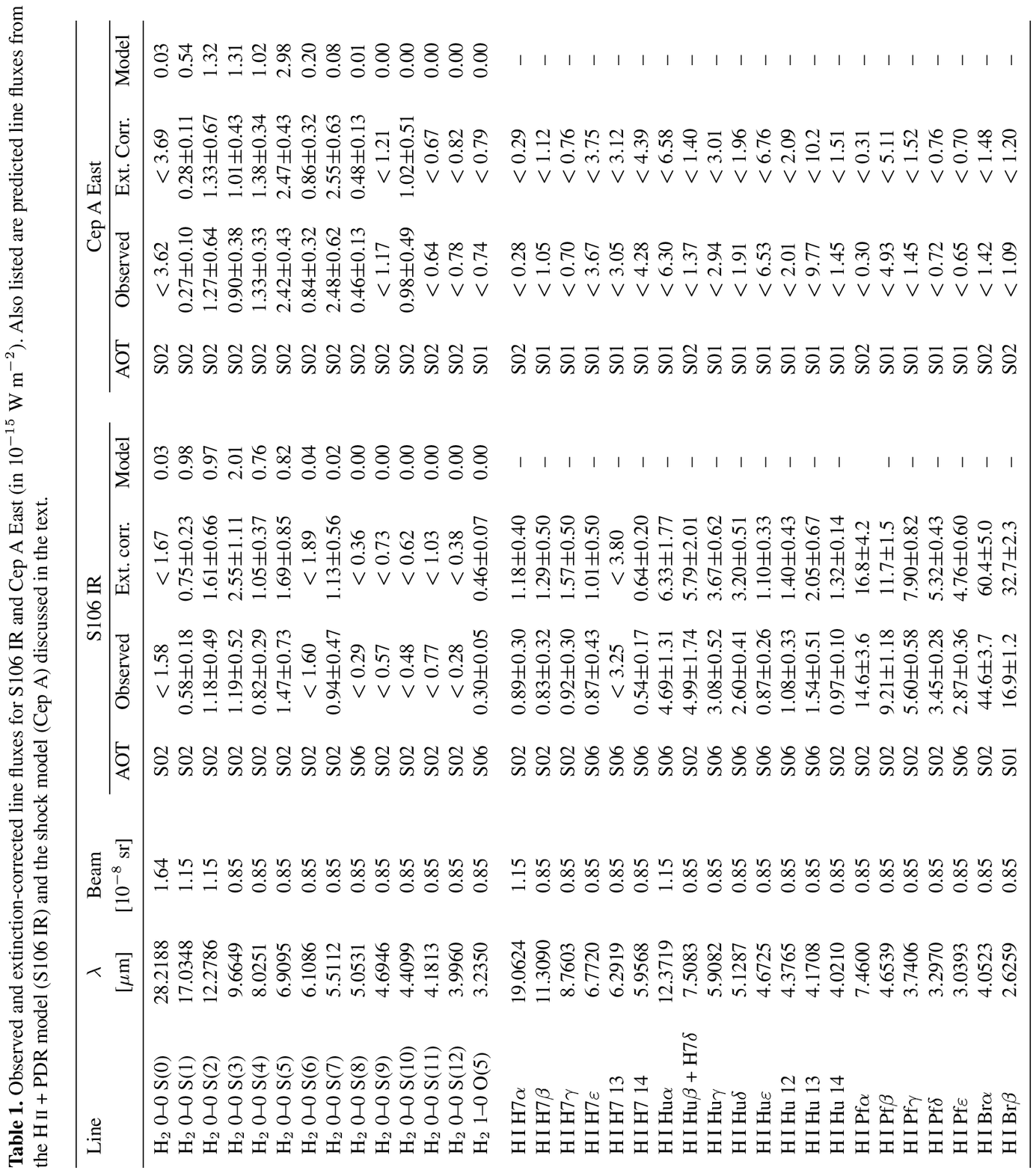,angle=180}}
\end{figure*}
\begin{figure*}
\vspace*{-5.5cm}
\hspace*{-0.5cm}
\centerline{\psfig{figure=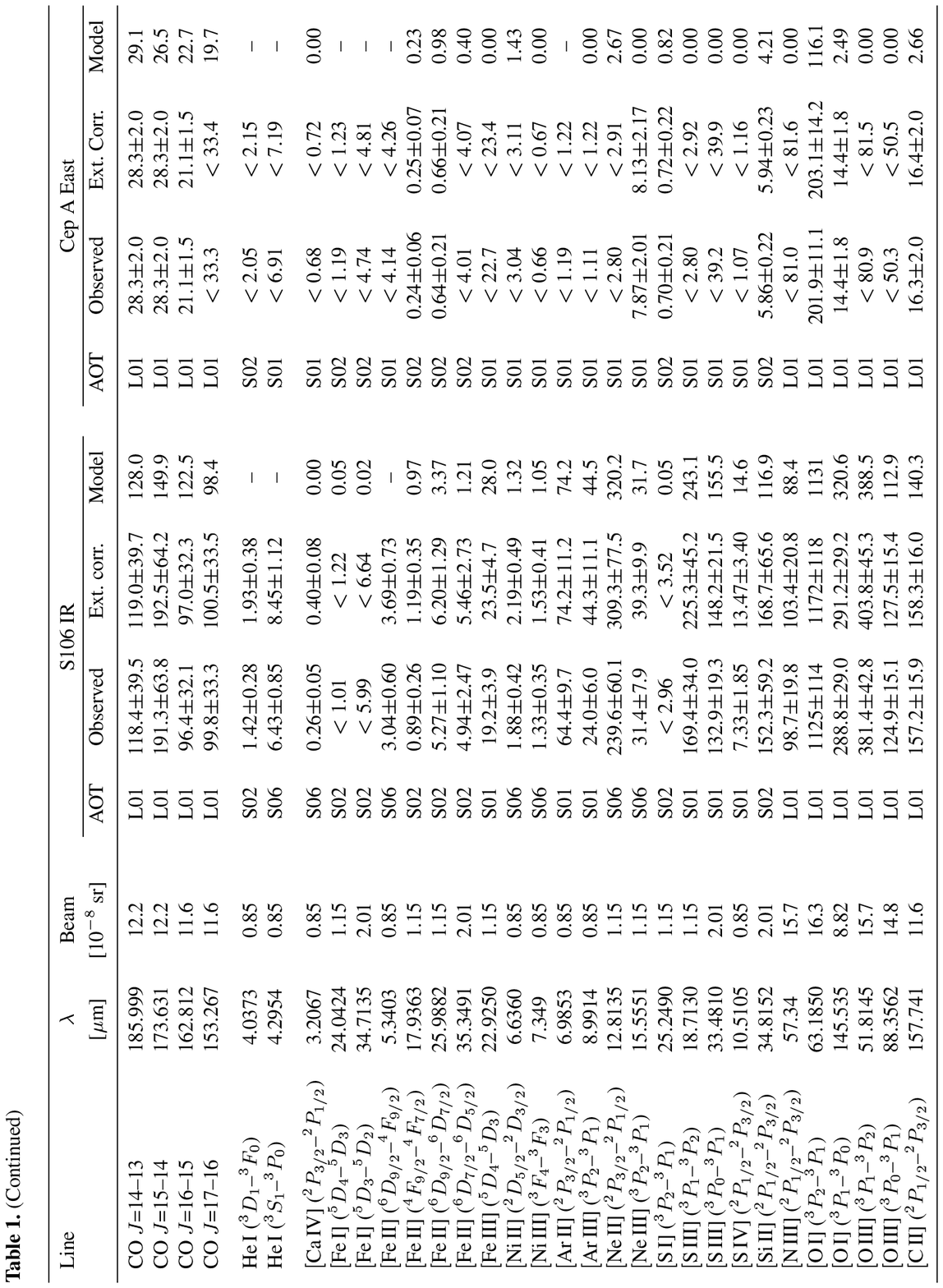,angle=180}}
\end{figure*}
\setcounter{table}{1}
ISO Short Wavelength (2.4--45~$\mu$m) Spectrometer (SWS; de Graauw 
et al. 1996) full grating scans (``AOT S01'') of S106 IR and Cep~A East were 
obtained in ISO revolutions 335 (JD~2450373.740) and 843 (JD~2450880.050), 
respectively. Each of these observations took 2124 seconds of observing 
time. In addition to this, deeper SWS grating scans on selected molecular 
and fine transition lines in the two objects (``AOT S02'') were obtained in 
revolutions 134 (at JD~2450172.705; S106), 220 (JD~2450258.463; Cep~A) and 
566 (JD~2450603.706; Cep~A). In revolutions 558 (JD~2450596.103) and 580 
(JD~2450617.838), scans at the full SWS grating resolution (``AOT S06'') 
covering the wavelength ranges of 3.0--3.5, 4.0--6.8, 7.0--7.6 and 
12.1--16.5~$\mu$m were obtained for S106 as well. 
ISO Long Wavelength (43--197~$\mu$m) Spectrometer 
(LWS; Clegg et al. 1996) grating scans (``AOT L01'') of S106 IR and 
Cep~A East were obtained in revolutions 134 (JD~2450172.726) and 566 
(JD~2450603.681), respectively. Data were reduced in a standard fashion 
using calibration files corresponding to OLP version 7.0, after which they 
were corrected for remaining fringing and glitches. To increase the S/N 
in the final spectra, the detectors were aligned and statistical outliers were 
removed, after which the spectra were rebinned to a lower spectral resolution.

The SWS full grating scans consist of twelve different 
grating scans, each covering a small wavelength region, 
which were joined to form one single spectrum. Because of the 
variation of the diffraction limit of the telescope with wavelength, 
different SWS bands use apertures of different sizes. This is 
illustrated in Figs.~1 and 2, where we show their relative position 
and size, overlaid on K-band (2.2~$\mu$m) images of S106 and 
Cep~A East by Hodapp \& Rayner (1991) and Hodapp (1994). For a source 
that is not point-like, one may see a discontinuity in the spectra   
at the wavelengths where a change in aperture occurs, which 
can indeed be seen in both the S106 and Cep~A spectra. The relative 
discontinuities in S106 are close to the maximum possible values, 
indicating that the source is extended across the entire SWS aperture 
at the longer wavelengths (33\arcsec $\times$ 20\arcsec). In Cep~A 
the discontinuities are smaller, pointing to a smaller size of the 
far-infrared source.

Since both spectrometers on board ISO use entrance apertures that are 
smaller than the beam size, a wavelength dependent correction has to be 
applied to the standard flux calibration when observing extended sources. 
We determined these correction factors by convolving the K-band images of 
S106 and Cep~A East, with the beam profile and applied these to the data 
before further analysis. The maximum correction to the flux was 8\% in 
the region around 40~$\mu$m. If the sources are more extended at the 
longer wavelengths or in specific lines than the K-band images we employed 
in estimating the diffraction losses, we will have underestimated the flux. 
The maximum error this could introduce in the flux calibration is 
$\approx$15\%.
\begin{figure*}
\centerline{\psfig{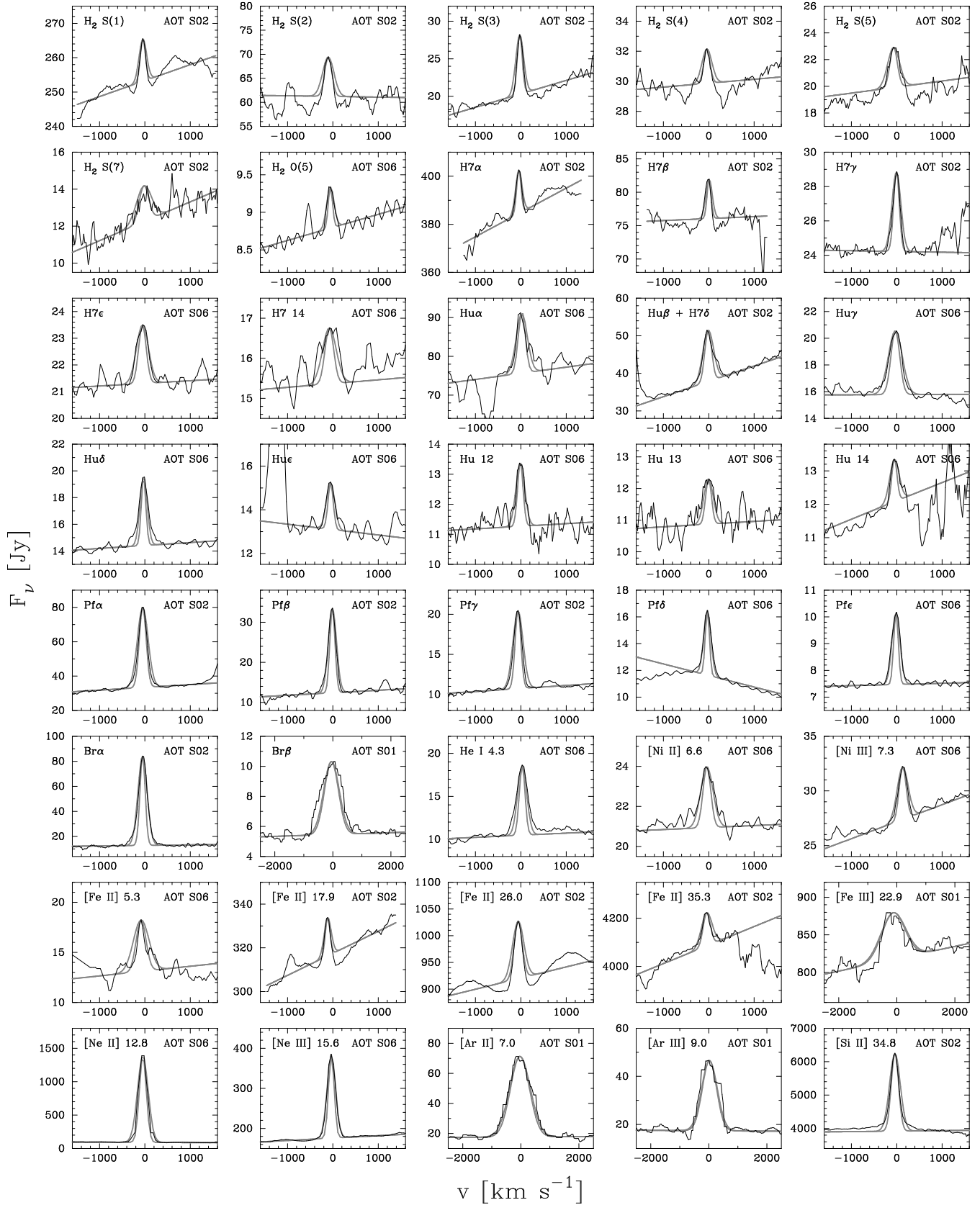}}
\caption[]{Detected lines in S106 IR. The velocities are 
heliocentric. The grey lines show the instrumental profiles for a 
point source (narrower profile) and an extended source (wide profile) 
filling the entire aperture.}
\end{figure*}
\setcounter{figure}{4}
\begin{figure*}
\centerline{\psfig{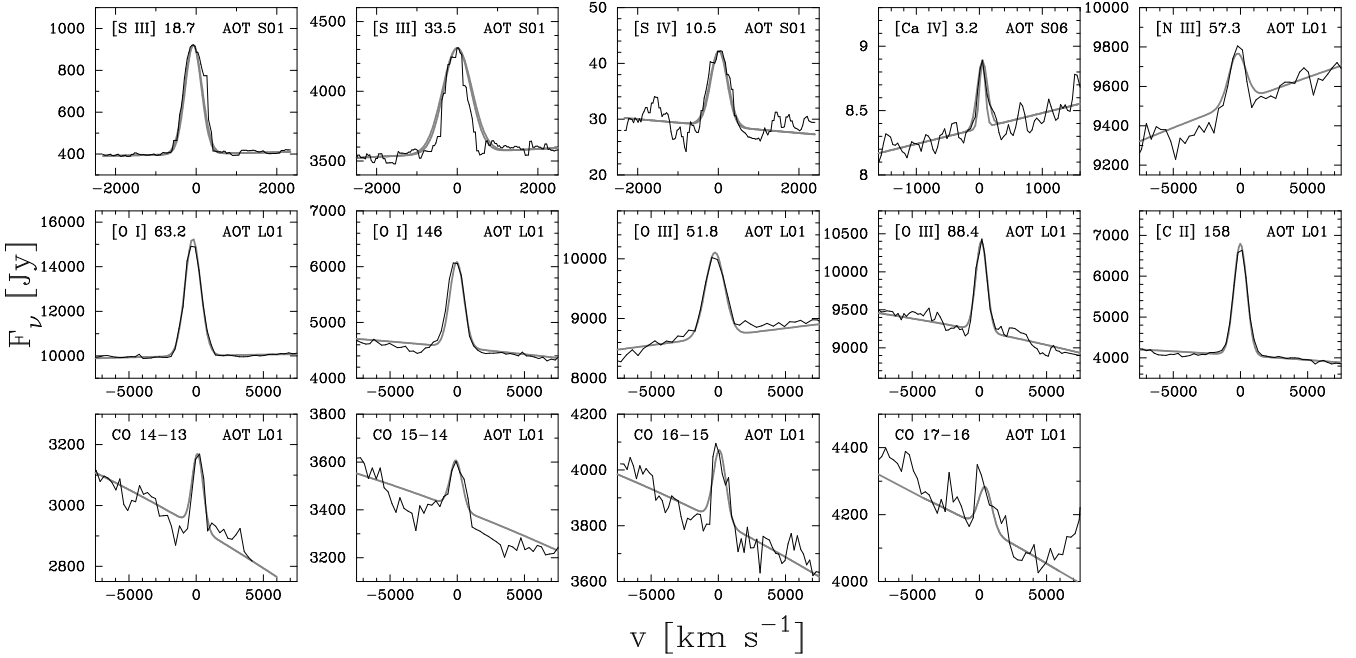}}
\caption[]{(Continued)}
\end{figure*}
\begin{figure*}
\centerline{\psfig{figure=lines_cepa.ps,width=17.8cm}}
\caption[]{Same as Fig.~5 for Cep~A East.}
\end{figure*}

Figures~3 and 4 show the resulting SWS and LWS full grating scans for 
S106 IR and Cep~A East. Scanned lines and measured line fluxes or upper 
limits (total flux for line with peak flux 3$\sigma$) are listed in 
Table~1. The errors listed in Table~1 include the errors in the 
absolute flux calibration; errors in line ratios used in the 
remainder of this paper may be lower. Plots of all detected lines, 
rebinned to a resolution $\lambda/\Delta\lambda$ of 1500 with an 
oversampling factor of four (SWS), or to a resolution of 500 (LWS), 
are given in Figs.~5 and 6. In the same figures we also show the 
line profiles expected for an unresolved point-source and an 
unresolved extended source filling the entire aperture. All detected 
lines in both S106 and Cep~A are compatible with these profiles. 
For lines measured with the full SWS grating resolution (AOT S02 or 
S06), this indicates that they have a FWHM of less than 
$\approx$ 150~km~s$^{-1}$.
\begin{table*}
\caption[]{Column densities of solid state absorption features
           towards S106 and Cep~A.}
\begin{flushleft}
\begin{tabular}{lcccccccc}
\hline\noalign{\smallskip}
Species        & $\lambda$ & $A_m$ &\,& \multicolumn{2}{c}{Cep~A East} &\,&
                                          \multicolumn{2}{c}{S106 IR}\\
\noalign{\vspace{0.02cm}}
\cline{5-6}\cline{8-9}\noalign{\vspace{0.05cm}}
 & [$\mu$m] & [cm molec$^{-1}$] && $\int \tau(\nu) d\nu$ [cm$^{-1}$] & $N$ [cm$^{-2}$] && 
                                   $\int \tau(\nu) d\nu$ [cm$^{-1}$] & $N$ [cm$^{-2}$]\\
\noalign{\smallskip}
\hline\noalign{\smallskip}
H$_2$O  & ~3.0 & 2.0 $\times$ 10$^{-16}$ && 1051 & 5.3 $\times$ 10$^{18}$ && 10.1    & 5.1 $\times$ 10$^{16}$\\
H$_2$O  & ~6.0 & 1.2 $\times$ 10$^{-17}$ &&   60 & 5.0 $\times$ 10$^{18}$ && $<$ 4.9 & $<$ 4.1 $\times$ 10$^{17}$\\
CO      & 4.67 & 1.1 $\times$ 10$^{-17}$ &&  3.8 & 3.5 $\times$ 10$^{17}$ && $<$ 1.9 & $<$ 1.7 $\times$ 10$^{17}$\\
CO$_2$  & 4.26 & 7.6 $\times$ 10$^{-17}$ &&   58 & 7.6 $\times$ 10$^{17}$ && $<$ 6.4 & $<$ 8.4 $\times$ 10$^{16}$\\
CO$_2$  & 15.2 & 1.1 $\times$ 10$^{-17}$ &&  9.2 & 8.4 $\times$ 10$^{17}$ && $<$ 1.5 & $<$ 1.3 $\times$ 10$^{17}$\\
CH$_4$  & 7.67 & 7.3 $\times$ 10$^{-18}$ &&  1.7 & 2.3 $\times$ 10$^{17}$ && $<$ 2.8 & $<$ 3.9 $\times$ 10$^{17}$\\
Silicate& ~9.7 & 1.2 $\times$ 10$^{-16}$ &&  965 & 8.0 $\times$ 10$^{18}$ &&  106    & 8.8 $\times$ 10$^{17}$\\
\noalign{\smallskip}
\hline
\end{tabular}
\end{flushleft}
\end{table*}

\section{Solid-state features}
The SWS spectrum of S106 (Fig.~3) consists of a relatively smooth 
continuum, with numerous strong emission lines superimposed. The 
continuum appears to consist of two components, with the break 
between the two occurring around 10~$\mu$m. This is similar to 
the situation seen in other regions of recent star formation, where 
these two components are commonly ascribed to emission due to dust 
close to the star (2--10~$\mu$m) and small dust grains in the 
wide circumstellar environment (10--200~$\mu$m; e.g. 
Cesarsky et al. 1996). 

The familiar UIR bands at 3.3, 3.4, 6.2, 7.6, 7.8, 8.6, 11.3 and 
12.7~$\mu$m, usually attributed to polycyclic aromatic hydrocarbons 
(PAHs), are present in emission and strong. As first demonstrated 
by ground-based spatially resolved CVF spectroscopy of the 
3.3~$\mu$m band by Felli et al. (1984), these features do not 
come from the central point-source IRS4, but originate in the 
wide circumstellar environment. The band-strengths 
of the UIR features in our ISO spectra appear consistent with 
the ground-based measurements present in literature (Hefele 
\& H\"olzle 1980; Felli et al. 1984; Geballe et al. 1985).

An absorption band around 3.0~$\mu$m due to the O--H stretch mode 
of H$_2$O ice, previously undetected in S106, is present in the SWS 
spectrum. Although partly filled in by emission from the 7.6, 7.8, 8.6 
and 11.3~$\mu$m UIR bands, a 9.7~$\mu$m absorption feature due to the 
Si--O stretching mode in amorphous silicates also appears present 
in Fig.~3. The combination of the contamination by the UIR bands 
and the limited baseline can explain why this feature has not been 
detected from the ground. A weak dip around 18--19~$\mu$m in our 
ISO data may be identified with the Si--O bending mode of amorphous 
silicate, confirming the presence of silicate absorption in S106.

Since extinction in 
the continuum surrounding the 9.7~$\mu$m feature is small compared to 
the extinction within this feature, the extinction $A_\lambda$ at 
wavelength $\lambda$ across a non-saturated 9.7~$\mu$m feature can simply 
be obtained from the relation $A_\lambda$ = $-2.5 \log (I/I_0)$. Using an 
average interstellar extinction law which includes the silicate feature 
(Fluks et al. 1994), we can then convert these values of $A_\lambda$ to 
a visual extinction, resulting in a value of $A_V$ = 13\fm7 toward 
S106~IR. This value is in excellent agreement with that of 
13\fm4 $\pm$ 2.7, derived towards stars in the S106 embedded cluster 
(Hodapp \& Rayner 1991), or the value of 15\mag $\pm$ 3\mag 
predicted from infrared H\,{\sc i} lines by Alonso-Kosta \& Kwan 
(1989). It is also in agreement with the extinction of 
$A_V$ = 12\mag\ towards the northern lobe of the S106 nebula 
(Felli et al. 1984). It is somewhat larger than that towards the 
southern lobe ($A_V$ = 8\mag; Felli et al. 1984) and much smaller 
than that towards the central source ($A_V$ = 21\mag; Eiroa et al. 1979).

For Cep~A East the SWS spectrum (Fig.~4) is dominated by absorption 
bands from a variety of ices. The O--H bend and stretch modes of 
H$_2$O at 3.0 and 6.0~$\mu$m are strong, as are the 4.27~$\mu$m 
$^{12}$C=O stretch and the 15.3~$\mu$m O=C=O bending mode of CO$_2$. 
The $^{12}$C=O stretch of CO at 4.67~$\mu$m is clearly detected. 
The absorption band around 7~$\mu$m, attributed to solid CH$_4$ 
(Boogert et al. 1996, 1997) is present as well. 
The unidentified absorption feature around 6.8~$\mu$m, also observed 
toward NGC~7538 IRS9 and RAFGL 7009S (Schutte et al. 1996; d'Hendecourt 
et al. 1996) is also present in Cep~A. The 9.7~$\mu$m amorphous 
silicate feature consists of a very deep, saturated, absorption. 
From the non-saturated wings of this feature we can again derive 
a visual extinction, resulting in a value of $A_V$ = 270\mag~ for 
Cep~A East. This value is within the range of $A_V$ = 75--1000\mag~ 
extinction for the central source and nebula reported by 
Lenzen et al. (1984). PAHs appear absent in Cep~A East.

From the integrated optical depth $\int \tau(\nu) d\nu$ of a non-saturated 
absorption feature we can compute a column density $N$ using an 
intrinsic band strength $A_m$. For H$_2$O, CO, CO$_2$ and CH$_4$ ices, 
values of $A_m$ were measured by Gerakines et al. (1995) and 
Boogert et al. (1997). For silicates, $A_m$ is taken from 
Tielens \& Allamandola (1987), based on the lab measurements by Day (1979). 
Integrated optical depth and column density values are listed in 
Table~2. The derived abundances of 100:6:15 for H$_2$O:CO:CO$_2$ 
in Cep~A East are within the range of values observed in other 
lines of sight (Whittet et al. 1996; d'Hendecourt et al. 1996).

\section{Hydrogen recombination lines}
A rich spectrum of H\,{\sc i} recombination lines is present 
in the SWS data of S106 (Table~1). In contrast, H\,{\sc i} 
lines are absent from the infrared spectra of Cep~A East. 
It is not clear whether the H\,{\sc i} lines observed with 
the rather large (27\arcsec $\times$ 14\arcsec) SWS beam
in S106 are dominated by the strong ($\dot{M}$ $\simeq$ 
10$^{-5}$~M$_\odot$~yr$^{-1}$) stellar wind (Hippelein \& 
M\"unch 1981; Felli et al. 1984), or could be due to the 
extended H\,{\sc ii} region seen in H$\alpha$ and Br$\gamma$ 
(Bally et al. 1998; Hayashi et al. 1990; Greene \& Rayner 1994; 
Maillard et al. 1999).

A clue to what is the case could come from the observed line 
profiles. As can be seen from Fig.~5, the H\,{\sc i} lines 
observed in S106 are wider than the instrumental line profile 
for a point source. Unfortunately, the effect of line broadening 
of a compact source with FWHM of a few hundred km~s$^{-1}$, 
as expected for the stellar wind (Felli et al. 1985; 
Drew et al. 1993), would be about the same as that of 
observing narrow recombination lines from an extended 
H\,{\sc ii} region. However, from the fact that our 
Br$\alpha$ line flux of 1.7 $\times$ 10$^{-14}$~W~m$^{-2}$ 
is identical to that observed by other authors in 
smaller apertures (Garden \& Geballe 1986; 
Persson et al. 1988; Drew et al. 1993), we conclude 
that at least for Br$\alpha$ the dominant component 
must be the compact wind. In view of the limited 
spectral resolution of our data, a full analysis of the 
wind structure is beyond the scope of the present paper.

\section{Molecular hydrogen emission}
Both S106 and Cep~A East are well-known sources of extended molecular 
hydrogen emission (Longmore et al. 1986; Hayashi et al. 1990; 
Bally \& Lane 1982; Doyon \& Nadeau 1988; Goetz et al. 1998). All 
ground-based H$_2$ measurements in literature refer to the 
ro-vibrational transitions, with upper energy levels $E((\rm J)/k$ 
above 5000~K. These do not probe the bulk of the H$_2$, which is 
expected to be at much lower temperatures. However, in both 
S106 and Cep~A East we have detected many pure rotational 
lines of H$_2$ (Table~1), which have much lower upper energy 
levels and are therefore able to directly probe the physical 
conditions in the dominant chemical species. And since 
their transition probabilities are quite small, these lines are
optically thin and the excitation temperature will be close to 
the kinetic temperature of the gas.

From the H$_2$ line fluxes $I(J)$ listed in Table~1 it is possible to 
calculate the apparent column densities of molecular hydrogen in the 
upper J levels, averaged over the SWS beam, $N(J)$, using 
$N(J) = \frac{4\pi I(J)}{A} \frac{\lambda}{hc}$, with $\lambda$ the 
wavelength, $h$ Planck's constant and $c$ the speed of light. The 
transition probabilities $A$ were taken from Turner et al. (1977). 
Line fluxes were corrected for extinction using the average 
interstellar extinction law by Fluks et al. (1994). For S106 IR we 
adopted the value of $A_V$ = 13\fm7 derived in Section~3. 
The fact that the 0--0 S(3) line of H$_2$, with a wavelength near the 
center of the 9.7~$\mu$m amorphous silicate feature, was detected 
in Cep~A indicates that it cannot suffer from the same extinction 
as the continuum; the H$_2$ emission must originate in a spatially 
separate region from the continuum. Therefore 
we adopt a much smaller value of $A_V$ = 2\mag, expected to be a 
reasonable value for an origin in either a shock or PDR, for 
the emission lines observed toward Cep~A East. Since the extinction 
correction at these mid-infrared wavelengths is small, a mis-estimate 
of $A_V$ by a few magnitudes will not affect our results.
\begin{figure}[t]
\centerline{\psfig{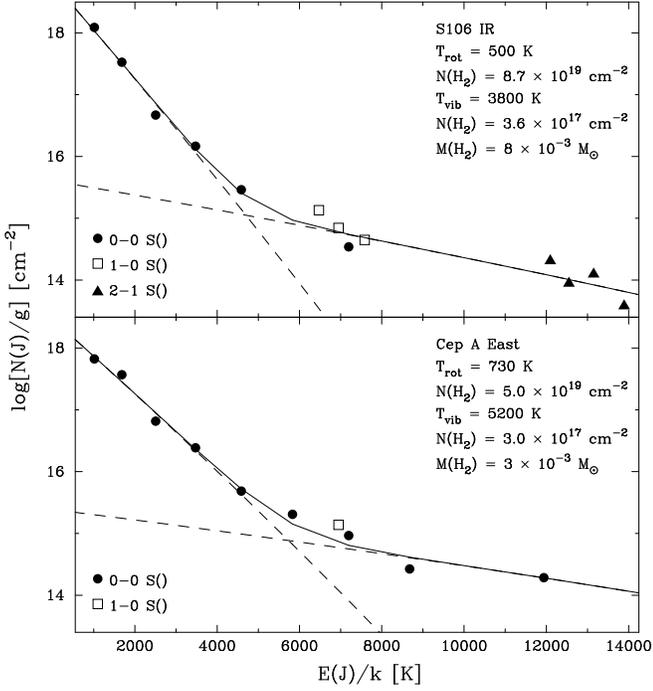}}
\caption[]{H$_2$ excitation diagrams for S106 (top) and Cep A East (bottom). 
ISO observations of pure rotational lines are indicated by the circles. 
Triangles and squares indicate ground-based measurements of ro-vibrational 
lines from literature. The dashed lines give the Boltzmann distribution fits 
to the low-lying pure rotational lines and all lines with upper level 
energies above 5000~K. The solid line shows the sum of both contributions.}
\end{figure}
\begin{figure}
\centerline{\psfig{figure=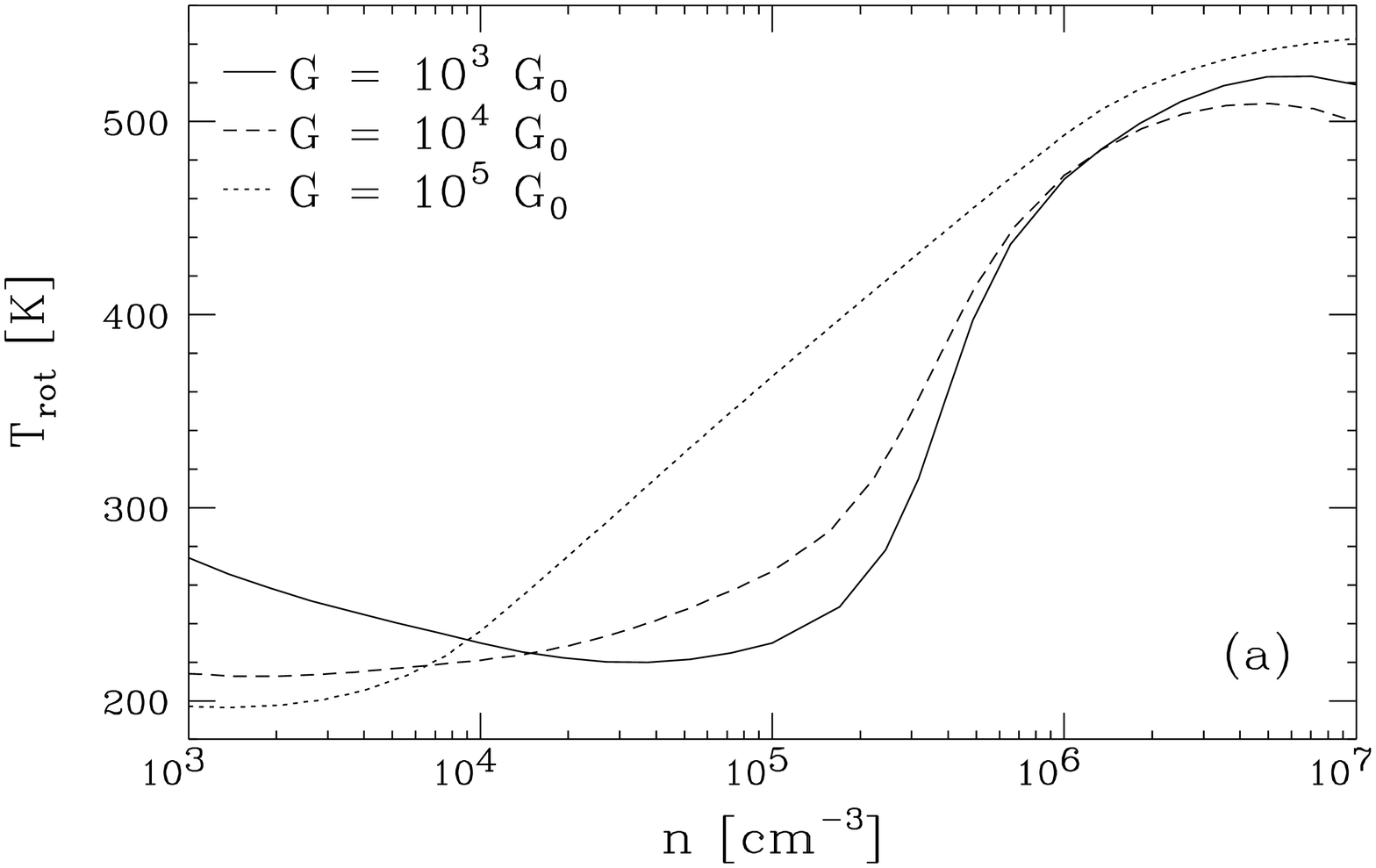,width=8.5cm,angle=90}}
\vspace*{0.2cm}
\centerline{\psfig{figure=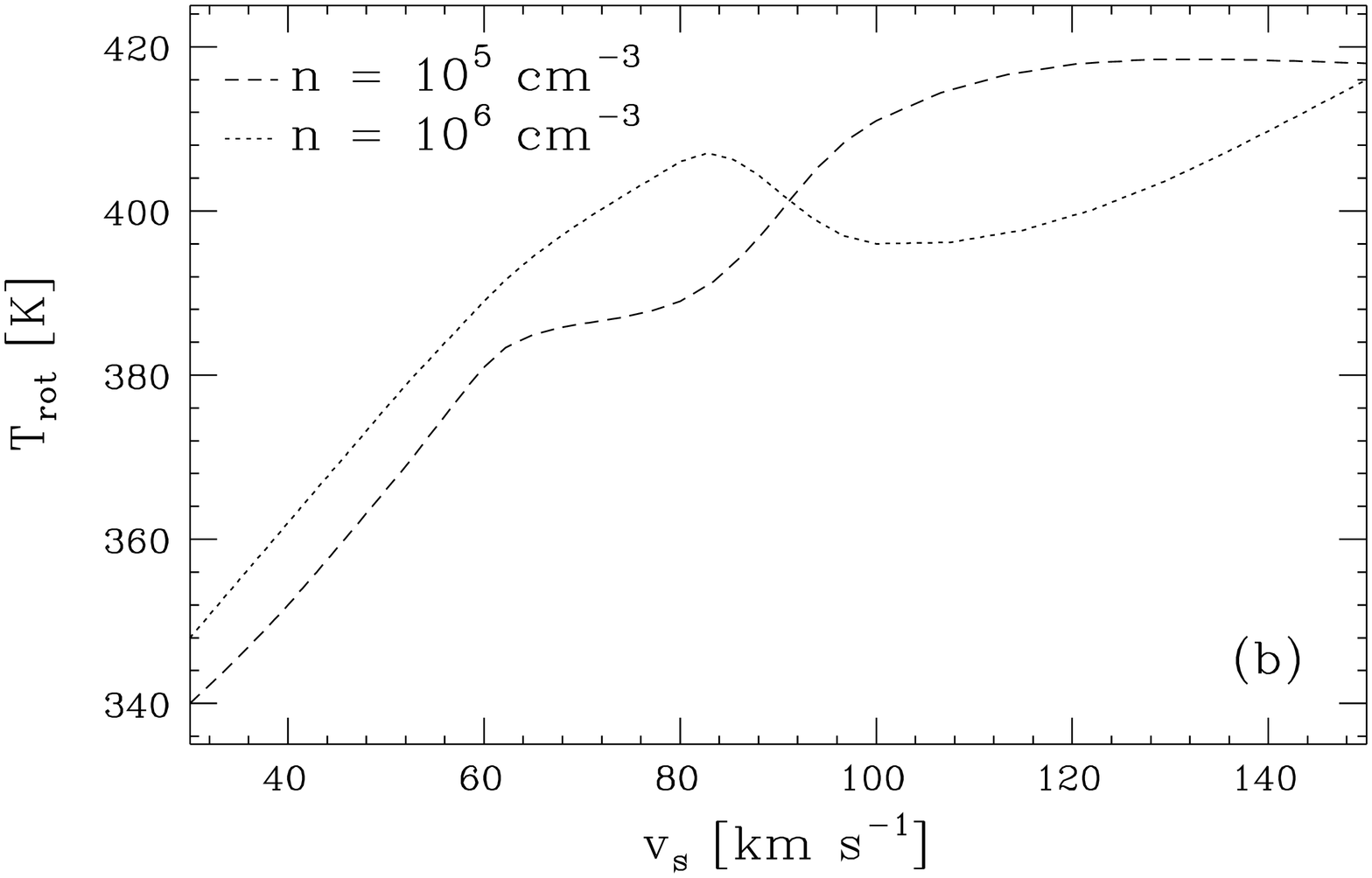,width=8.5cm,angle=90}}
\vspace*{0.2cm}
\centerline{\psfig{figure=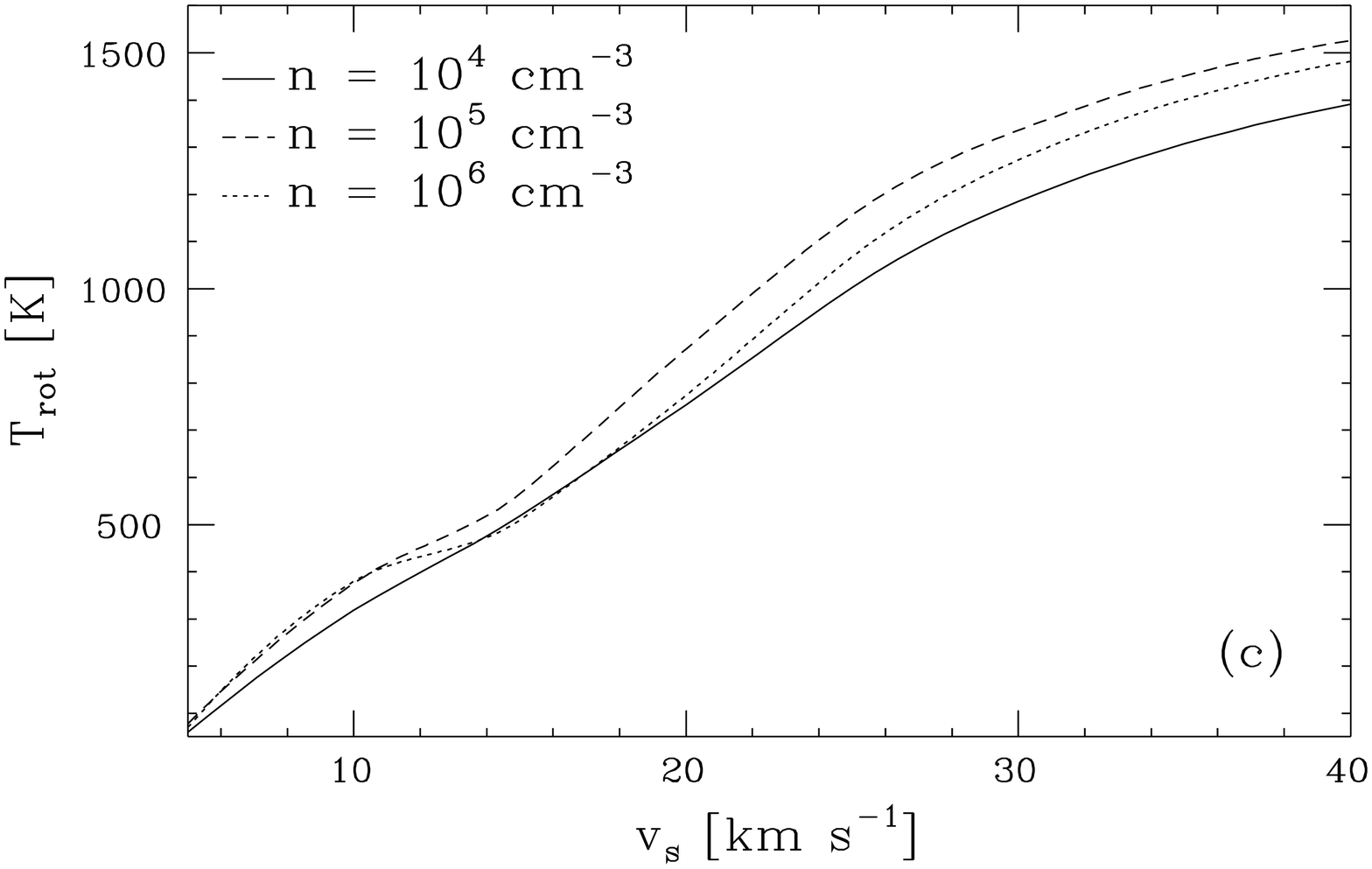,width=8.5cm,angle=90}}
\caption[]{Theoretical relation between a) $T_{\rm rot}$(H$_2$) and $n$ for 
PDR models, b) $T_{\rm rot}$(H$_2$) and $v_s$ for J-shock models, and c) 
$T_{\rm rot}$(H$_2$) and $v_s$ for C-shock models.}
\end{figure}

A useful representation of the H$_2$ data is to plot the log of
$N(\rm J)/g$, the apparent column density for a given J upper level 
divided by the statistical weight, versus the energy of the upper level, 
taken from Dabrowski (1984). These plots are shown in Fig.~7. Also 
plotted in these figures are several measurements of H$_2$ ro-vibrational 
lines in S106 and Cep~A from literature (Longmore et al. 1986; Tanaka et al. 
1989; Bally \& Lane 1982). Although these measurements were taken at the 
same positions on the sky as ours, the beam sizes used were different 
and hence one may expect to see systematic differences in the derived 
specific intensities if the source of H$_2$ radiation is more extended 
than the beam size. In fact only the S106 measurements by Longmore et al. 
(1986), taken with the smallest beam diameter, 12\arcsec, differ 
systematically from the other measurements, indicating that the extent 
of the H$_2$ emitting region in S106 is probably somewhat larger than 
their beam size. This agrees well with H$_2$ 1--0~S(1) images of the 
region (Hayashi et al. 1990; Maillard et al. 1999). 
In Fig.~7 the Longmore et al. measurements of S106 were scaled to the 
Tanaka et al. data.

The statistical weight $g$ used in Fig.~7 is a combination of the 
rotational and nuclear spin components. We have assumed the high temperature
equilibrium relative abundances of 3:1 for the ortho and para forms of
H$_2$ (Burton et al. 1992). For a Boltzmann distribution, the points in 
the plot shown in Fig.~7 should form a nearly straight line whose slope  
is inversely proportional to the excitation temperature, while 
its intercept is a measure of the total column density of warm gas. 
Using the formula by Parmar et al. (1991) and the rotational constants 
by Dabrowski (1984) to fit our data points in the low-lying pure 
rotational levels to a Boltzmann distribution, we arrive at values of 
500~K and 9 $\times$ 10$^{19}$~cm$^{-2}$ 
and 730~K and 5 $\times$ 10$^{19}$~cm$^{-2}$ for S106 IR and Cep~A East, 
respectively. Using the distances to S106 and Cep~A of 1200 and 690~pc 
(Rayner 1994; Mel'nikov et al. 1995), this corresponds to total 
molecular hydrogen masses of 0.008 and 0.003 M$_\odot$ within the SWS 
beam. Using the alternative distance estimate towards S106 of 600~pc 
(Staude et al. 1982) would decrease the S106 H$_2$ mass to 0.004 M$_\odot$.
The fitted Boltzmann distributions are shown as the leftmost dashed lines in 
Fig.~7. The fact that the points for the ortho and para form of H$_2$ 
lie on the same line proves that our assumption on their relative abundances 
is correct.

As can be seen from Fig.~7, both the pure-rotational and ro-vibrational 
lines of H$_2$ with upper level energies higher than 5000~K deviate 
significantly from the leftmost dashed line. In both cases a Boltzmann 
distribution was fitted to these lines as well, shown as the rightmost 
dashed line in both figures. In the case of S106, the relative location 
of the 1--0 and 2--1 S(1)-lines in Fig.~7 are indicative of fluorescent 
excitation through UV pumping (Draine \& Bertoldi 1996; Black \& 
van Dishoeck 1997). The fitted Boltzmann distribution to the higher energy 
levels has thus no physical meaning, but may still be useful to provide 
a simple parametrization of the relative population of the energy 
levels. In the case of Cep~A, the lines may indicate the presence of a 
smaller column (3 $\times$ 10$^{17}$~cm$^{-2}$) of hot (a few 1000~K) 
molecular hydrogen, in addition to the large column of warm gas.

Employing predictions of H$_2$ emission from PDR, J-shock and C-shock models 
by Burton et al. (1992), Hollenbach \& McKee (1989) and Kaufman \& Neufeld 
(1996), we determined the excitation temperature $T_{\rm rot}$ from 
the low-lying pure rotational levels from these models as a function of 
density $n$ and either incident FUV flux $G$ (in units of the average 
interstellar FUV field G$_0$ = 1.2 $\times$ 10$^{-4}$ 
erg~cm$^{-2}$~s$^{-1}$~sr$^{-1}$; Habing 1968) or shock velocity $v_s$ in 
an identical way as was done for the observations. The resulting relations 
between $T_{\rm rot}$ and $n$ or $v_s$ are shown in Fig.~8. As can be 
seen from these plots, the PDR and J-shock models predict a fairly small 
(200--540~K) range of resulting excitation temperatures, whereas in the 
C-shocks this range is much larger (100--1500~K). Furthermore, we see 
that in the model predictions for shocks the resulting $T_{\rm rot}$ 
does not depend much on density, whereas for PDRs it does not depend much 
on $G$, suggesting that once the mechanism of the H$_2$ emission is 
established, it can be used to constrain $v_s$ or $n$ in a straightforward 
way.

Comparing the excitation temperatures of 500 and 730~K for S106 and Cep~A 
with those plotted in Fig.~8, we note that for S106 this falls well within 
the range of PDR- and C-shock model predictions, but are too high compared 
to the ones expected from J-shocks. The observed bright PAH emission 
features (Sect.~3) and the atomic fine-structure line spectrum (Sect. 6) point 
towards a PDR origin for the H$_2$ emission in S106. The higher temperature 
for Cep~A can only be reproduced by the C-shock models. Therefore we 
tentatively conclude that a dense ($\ge$ 10$^6$~cm$^{-3}$) PDR seems to be 
the best candidate to explain the observed H$_2$ emission in S106 IR and 
a slow ($\approx$ 20~km~s$^{-1}$) non-dissociative shock can explain the 
observed warm column of H$_2$ in Cep~A East. Since the 
regions we are looking at probably only fill part of the SWS beam, the 
absolute intensity of the H$_2$ emission listed in Table~1 can also be 
reproduced by these same models by varying the beam filling factor.
\begin{figure}[t]
\centerline{\psfig{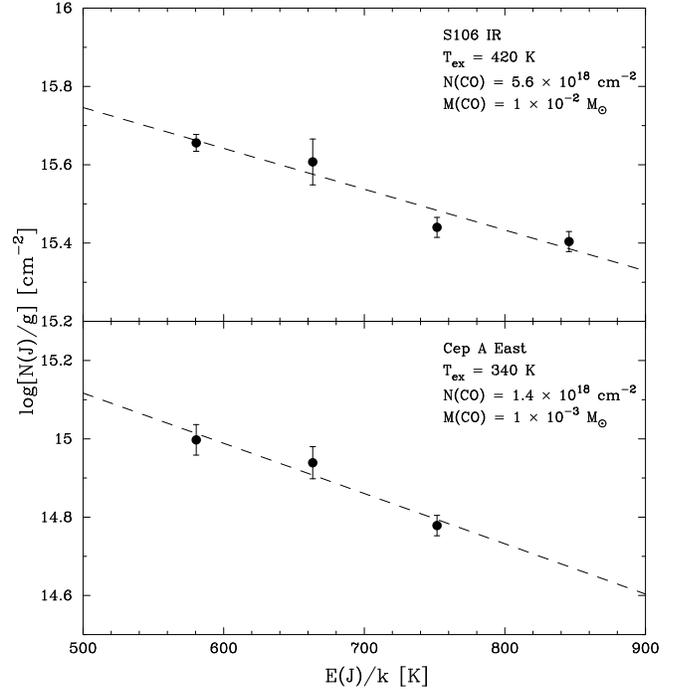}}
\caption[]{CO excitation diagrams for S106 (top) and Cep~A East (bottom). 
The dashed lines give the fits of the Boltzmann distribution to the data 
points.}
\end{figure}

\section{Carbon-monoxide emission lines}
In both S106 IR and Cep~A East several ro-vibrational emission lines 
due to gas-phase CO were detected in the long-wavelength part of the 
LWS spectra (Figs.~5 and 6). Similar to what was done for the H$_2$ 
emission in the previous section, we constructed a CO excitation 
diagrams, using molecular data from Kirby-Docken \& Liu (1978). 
They are shown in Fig.~9. The temperature and column of CO resulting 
from the Boltzmann fit to this excitation diagram are 420~K and 
5.6 $\times$ 10$^{18}$~cm$^{-2}$ and 340~K and 
1.4 $\times$ 10$^{18}$~cm$^{-2}$ for S106 and Cep~A East, respectively.
Corresponding CO masses are 1 $\times$ 10$^{-2}$~M$_\odot$ (S106 
at 1200~pc; for the 600~pc distance estimate this mass estimate 
would decrease to 3 $\times$ 10$^{-3}$~M$_\odot$) and 
1 $\times$ 10$^{-3}$~M$_\odot$ (Cep~A East).

The CO excitation temperature of 420~K for S106 is somewhat lower 
than that found from the H$_2$ lines, in agreement with what is 
expected from a PDR, in which the CO emission arises in deeper 
embedded regions than the H$_2$. We conclude that a PDR is the 
most likely candidate for the source of the gas-phase CO emission 
in S106. For Cep~A East the CO excitation temperature of 340~K is 
much lower than that found from the H$_2$ lines. This behaviour 
is hard to reconcile with an origin in the C-shock invoked to 
explain the H$_2$ observations. However, a J-shock, necessary to 
explain the ionic lines which we will discuss in the next section, 
may produce the observed CO emission while only emitting 
little warm H$_2$. Since the LWS CO observations were made with 
a much larger beam than the SWS H$_2$ observations, an alternative 
explanation may be that we are observing emission 
from two spatially distinct regions. The relatively large 
mass of warm CO as compared to that seen in H$_2$ may also 
easily be explained in this scenario.

The observed CO lines have critical densities of around 
10$^6$~cm$^{-3}$. Therefore the detection of these lines 
also implies densities of this order of magnitude or higher 
in the originating region. For the shock seen in Cep~A East 
this may not be unreasonable. However, assuming that these 
densities would exist in the entire S106 PDR would be implausible. 
If these CO lines do indeed arise in the large-scale environment 
of S106, the PDR must therefore have a clumpy structure (e.g. 
Burton et al. 1990). An alternative explanation of the presence 
of these lines, would be to identify the originating region with 
the surface of the extended disk-like structure surrounding S106 
which could act like a PDR. With the present data-set we cannot 
make a distinction between these two possibilities.
\begin{figure*}
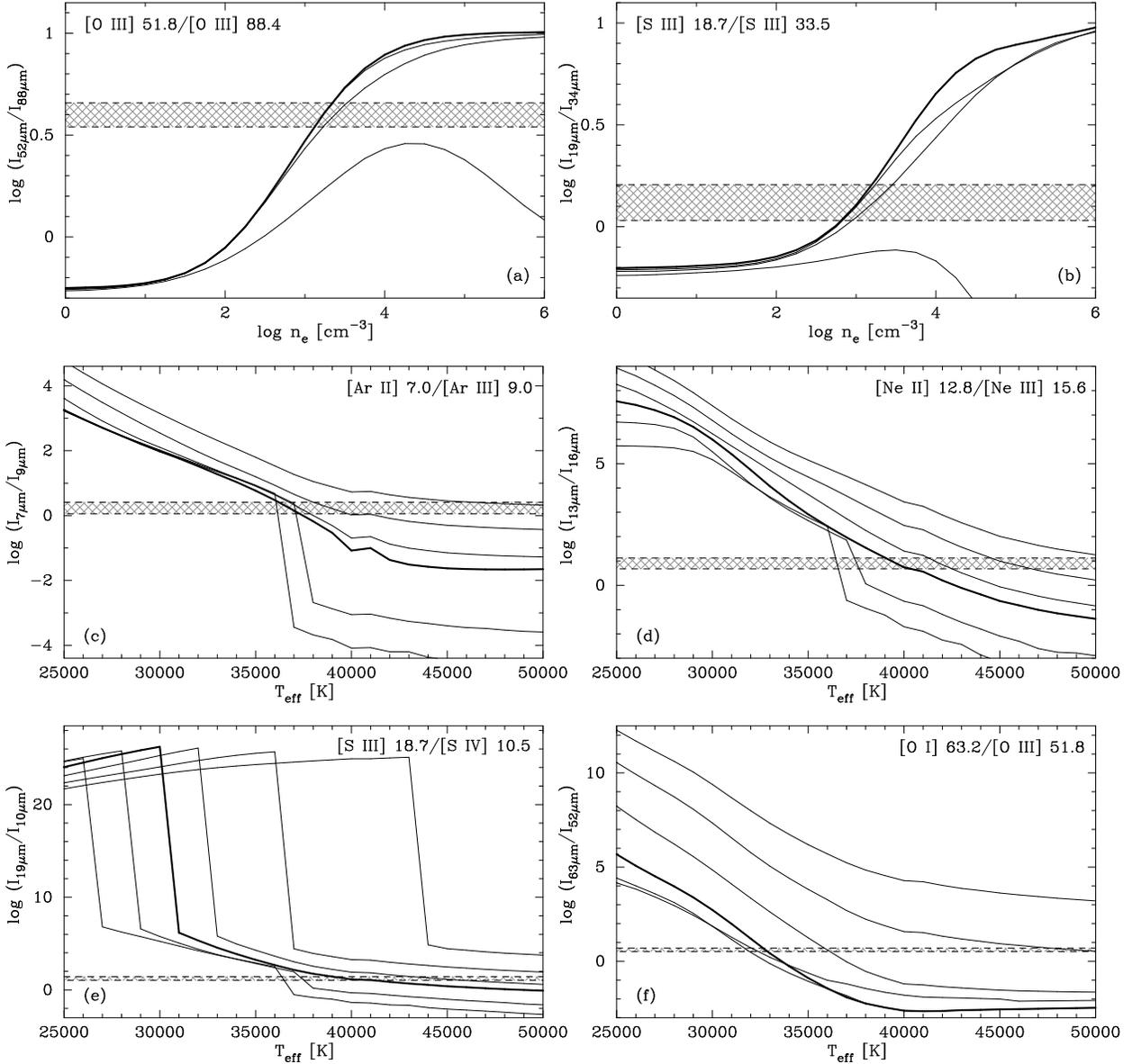

\centerline{\psfig{figure=oiiiratio.ps,width=8.0cm,angle=270}
            \hspace*{0.2cm}
            \hspace*{-0.3cm}
            \psfig{figure=siiiratio.ps,width=8.0cm,angle=270}
            \hspace*{0.2cm}}
\vspace*{0.3cm}
\centerline{\psfig{figure=ariiratio.ps,width=8.3cm,angle=270}
            \hspace*{-0.3cm}
            \psfig{figure=neiiratio.ps,width=8.3cm,angle=270}}
\vspace*{0.3cm}
\centerline{\psfig{figure=sivratio.ps,width=8.3cm,angle=270}
            \hspace*{-0.3cm}
            \psfig{figure=oiratio.ps,width=8.3cm,angle=270}}
\caption[]{{\sc cloudy} model predictions for various emission line ratios 
from the S106 H\,{\sc ii} region as function of electron density $n_e$ 
and temperature of the ionizing star. The hatched regions between the 
dashed lines show the error interval of the observed ratios in S106 IR.
The solid lines show the computed ratios for $\log n_e$ = 1.0--6.0, or 
$T_{\rm eff}$ = 15,000--50,000~K. The heavy curves indicate the lines 
with $\log n_e$ = 3.0 or $T_{\rm eff}$ = 40,000~K (best fit). 
(a) Behaviour of the [O\,{\sc iii}] 51.8~$\mu$m/[O\,{\sc iii}] 
88.4~$\mu$m line flux ratio. 
(b) The same for the [S\,{\sc iii}] 18.7~$\mu$m/[S\,{\sc iii}] 
33.5~$\mu$m ratio.
(c) Same for the [Ar\,{\sc ii}] 7.0~$\mu$m/[Ar\,{\sc iii}] 9.0~$\mu$m ratio. 
(d) Same for the [Ne\,{\sc ii}] 12.8~$\mu$m/[Ne\,{\sc iii}] 15.6~$\mu$m ratio. 
(e) Same for the [S\,{\sc iii}] 18.7~$\mu$m/[S\,{\sc iv}] 10.5~$\mu$m ratio. 
(f) Same for the [O\,{\sc i}] 63.2~$\mu$m/[O\,{\sc iii}] 51.8~$\mu$m ratio.} 
\end{figure*}

\section{Atomic fine structure lines}
Important constraints on the physical conditions in the line emitting 
region come from the observed fine structure lines. As a first 
step towards identifying the mechanism responsible for the observed emission 
we can look at the mere presence of certain lines. The observed lines 
with high ionization potentials in S106, such as [O\,{\sc iii}], 
[Ar\,{\sc iii}], [Ne\,{\sc iii}], [Ni\,{\sc iii}], [Fe\,{\sc iii}] and 
[S\,{\sc iv}] can only originate in the H\,{\sc ii} region surrounding 
S106. The PAH emission as well as the molecular lines observed towards 
S106 are indicative of the presence of a PDR as well. This PDR might 
contribute to the observed [Fe\,{\sc ii}], [Ni\,{\sc ii}], [Si\,{\sc ii}], 
[O\,{\sc i}] and [C\,{\sc ii}] emission. We thus need to model the 
fine-structure lines in S106 IR as arising in the combination of an 
H\,{\sc ii} model and a PDR.

We have used the photo-ionization code {\sc cloudy} (version 90.04; 
Ferland 1996) to generate model predictions for line strengths for the 
H\,{\sc ii} region surrounding S106 IR, assuming a spherical geometry 
and constant hydrogen density throughout the region. We generated a grid 
of models in which the input spectrum, taken from Kurucz (1991) models 
for a stellar photosphere, and the electron density in the H\,{\sc ii} 
region were varied. The total luminosity of the model was fixed to the 
total bolometric luminosity of S106 of 4.2 $\times$ 10$^4$~L$_\odot$ 
(computed from the spectral energy distribution using data from 
literature and assuming a distance of 1.2~kpc). The line ratios of the 
observed [O\,{\sc iii}] and [S\,{\sc iii}] fine structure lines are 
expected to depend mainly on density, and hardly on the temperature. 
Their behaviour in our H\,{\sc ii} region model together with the 
observed line ratios in S106 are shown in Fig.~10. From the fact that both  
line ratios agree on the thus obtained value for the electron density 
in the H\,{\sc ii} region, 1.3--2.5 $\times$ 10$^3$~cm$^{-3}$, we 
conclude that our assumption of a constant density is reasonable. 

Several other lines ratios ([Ar\,{\sc ii}]/[Ar\,{\sc iii}], 
[Ne\,{\sc ii}]/[Ne\,{\sc iii}] and [S\,{\sc iii}]/[S\,{\sc iv}]) 
depend mainly on the effective temperature of the star and 
do not depend on density much. Their behaviour in the {\sc cloudy} 
models for S106 is shown in Fig.~10 as well. From these plots, it 
can be seen that the source of ionizing photons should have a 
temperature between 37,000 and 40,000~K, corresponding to a spectral 
type of O6--O8, in good agreement with earlier determinations 
(Gehrz et al. 1982; Staude et al. 1982). 
However, not all observed line ratios yield the same 
temperature for the central star. Most likely this is due to the 
fact that the Kurucz models used for the input spectrum do not 
include the opacity shortward of the He\,{\sc ii} ionization limit 
due to the stellar wind, which is expected to be strong in the case 
of S106. The true temperature for the central source is therefore 
expected to be around the lower range of temperatures deduced from 
the Kurucz synthetic photospheres, i.e. around 37,000~K. For 
illustrative purposes, we also show the ratio of 
[O\,{\sc i}] 63.2~$\mu$m/[O\,{\sc iii}] 51.8~$\mu$m for our model 
H\,{\sc ii} region, showing that virtually all the atomic oxygen 
will be in the form of [O\,{\sc iii}] and that we can thus safely 
attribute all of the observed [O\,{\sc i}] emission to the PDR.

To be able to explain the observed intensity of 
1.3 $\times$ 10$^{-2}$~erg~s$^{-1}$~cm$^{-1}$~sr$^{-1}$ for the 
[Si\,{\sc ii}] line at 34.8~$\mu$m in S106, the Tielens \& Hollenbach 
(1985) PDR models require a density higher than $\approx$ 10$^5$~cm$^{-3}$ 
and $G$ $\ge$ 10$^5$~G$_0$. This regime can also reproduce the observed 
ratios of [Si\,{\sc ii}] and the [Fe\,{\sc ii}] and [O\,{\sc i}] lines, 
although to reproduce both 
the exact strength of the [Si\,{\sc ii}] emission and the relative strength 
to that to the [Fe\,{\sc ii}] lines requires a [Fe\,{\sc ii}] depletion of 
$\approx$~30\% higher than the one assumed in the Tielens \& Hollenbach models. 
An alternative explanation could be that the PDR doesn't fill the SWS 
aperture at 35 micrometer (33\arcsec $\times$ 20\arcsec) completely, 
increasing the surface brightness of all lines. The predicted 
intensities of [S\,{\sc i}] and [Fe\,{\sc i}] are sufficiently low to be 
undetectable, in agreement with the observations.

Towards Cep A East, only fine structure lines that can be produced in shocks 
were observed. In contrast to both C- and J-shocks, PDRs do not produce 
significant quantities of [S\,{\sc i}] emission (Tielens \& Hollenbach 1985). 
Therefore this line must be completely due to one or 
more shocks. C-shocks only contain trace fractions of ions and hence cannot 
explain the observed [Fe\,{\sc ii}], [Si\,{\sc ii}] and [C\,{\sc ii}] 
emission. Hence the simplest hypothesis would be to try to explain the 
observed fine-structure lines in Cep~A
in terms of a J-shock model. To explain the [S\,{\sc i}] surface brightness 
of 1.0 $\times$ 10$^{-3}$~erg~s$^{-1}$ cm$^{-2}$~sr$^{-1}$ requires a 
moderately dense (10$^4$--10$^5$ cm$^{-3}$) pre-shock gas. If the shock 
does not fill the ISO beam completely, as is expected, this number will 
go up. The absence of [Fe\,{\sc i}] emission does indicate that we are 
dealing with higher (10$^5$--10$^6$~cm$^{-3}$) densities, so line ratios 
seem to provide more reliable constraints in this case. However, the aperture 
sizes for lines measured at different wavelengths are also different in 
some cases, making this method only reliably applicable to line ratios 
measured in identical SWS or LWS apertures. Therefore the ratio of  
[Fe\,{\sc ii}] 26.0~$\mu$m to [S\,{\sc i}] 25.2~$\mu$m might provide the 
most reliable indicator of physical conditions in the shock. To reproduce 
the observed value of 0.26 with the Hollenbach \& McKee (1989) J-shock models 
requires either moderately dense 
(10$^5$~cm$^{-3}$) gas with a low (30~km~s$^{-1}$) shock velocity or 
a high (10$^6$~cm$^{-3}$) density with a faster ($\approx$ 60~km~s$^{-1}$) 
shock. To have the predicted [Fe\,{\sc i}] emission sufficiently low 
to explain our non-detections of this species, the moderately dense, 
slow J-shock model is required. This regime also reproduces the 
relative [Si\,{\sc ii}] strength.

As was discussed in the previous section, one or multiple J-shocks cannot 
reproduce the observed H$_2$ emission (although a contribution to this can be 
expected). The success of the J-shock model in explaining the observed 
fine-structure lines and the absence of PAH emission in Cep~A East, 
excluding the possibility of a significant contribution from a PDR, leads 
us to pose that a combination of one or more J- and C-shocks must be 
responsible for the observed emission in Cep~A. The presence of more than 
one type of shock in the region could be linked to the reported multiple 
episodes of outflow activity from the embedded source (Narayanan \& Walker 
1996). In the presence of both a J- and a C-shock, the [Si\,{\sc ii}] and 
[Fe\,{\sc ii}] emission would originate completely in the J-shock, whereas 
both the C- and the J-shock would contribute to the observed [S\,{\sc i}] 
and H$_2$ spectra. With the J-shock parameters derived from the 
fine-structure lines, an additional C-shock component with a shock-velocity 
of about 20~km~s$^{-1}$ is required to reproduce the observed pure 
rotational H$_2$ emission. The observed hot column of H$_2$ in Cep~A may 
then be due to formation pumping (the effect that H$_2$ gets re-formed 
with non-zero energy in the post-shock gas after being dissociated in the 
shock front) in the J-shock. To have the ro-vibrational lines of comparable 
intensity as the rotational lines (through collisional excitation) requires 
pre-shock densities of the order of 10$^6$~cm$^{-3}$.
For both S106 IR and Cep~A, predicted lines strengths of the best fit models  
are also listed in Table~1.

\section{Discussion and conclusions}
In the previous sections we saw that the infrared emission-line spectrum 
of S106 could be well explained by the presence of an H\,{\sc ii} region 
and a clumpy PDR. The average density in the H\,{\sc ii} region should 
be around 2 $\times$ 10$^3$~cm$^{-3}$. The density in the PDR, 
which was estimated to be 10$^5$--10$^6$~cm$^{-3}$, is several orders 
of magnitude higher, in agreement with one would expect in a scenario in 
which a young massive star has photo-ionized its natal cloud in all 
but the densest clumps.

In Section~7 we concluded that the incident FUV flux on the S106 
PDR should be rather high ($\ge$ 10$^5$~G$_0$). We also showed that 
the central source in S106 is of spectral type O8, with a total 
luminosity of 4.2 $\times$ 10$^4$~L$_\odot$. From a Kurucz (1991) 
model for a stellar photosphere with $T_{\rm eff}$ = 37,000~K 
and $\log g$ = 4.0, we compute that such a star emits a total FUV 
(6--13.6~eV) flux of 3.6 $\times$ 10$^{36}$~erg~s$^{-1}$~sr$^{-1}$. 
To dilute this stellar FUV field to a value of 10$^5$--10$^6$~G$_0$, 
the PDR must be at a location 3--10 $\times$ 10$^3$~AU away from the 
central star, corresponding to a projected distance of 5--17\arcsec. 
This projected distance is independent of the assumed distance 
towards S106. It is within the range allowed by the 
SWS entrance apertures and is compatible with an origin in the 
$\approx$ 30\arcsec~ diameter region of H$_2$ emission in the H$_2$ 
1--0 S(1) image of S106 by Hayashi et al. (1990). We conclude that 
a central O8 star is sufficient to produce the FUV 
radiation field reaching the S106 PDR.

These results agree well with other 
determinations. Schneider et al. (2000) have spatially resolved 
the 157.7~$\mu$m [C\,{\sc ii}] and 63.2~$\mu$m [O\,{\sc i}] 
emission in S106 and concluded that there is a PDR region 
with density $\simeq$~10$^6$~cm$^{-3}$ and strong (up to 
8 $\times$ 10$^5$~G$_0$) UV field in the immediate environment 
of S106 and a more extended dense (10$^5$--10$^6$~G$_0$) 
PDR region exposed to a lower intensity (300--500~G$_0$) 
UV field. We are clearly dominated by their first component. 
This same component may be identified with the high-density 
(10$^5$~cm$^{-3}$) fluorescent H$_2$ regions identified by 
Hayashi et al. (1990).

Both the presence of [S\,{\sc i}] 25.2~$\mu$m emission and the 
H$_2$ spectrum point unambiguously to shocked gas as the origin 
of the infrared emission lines in Cep~A East. A combination 
of a slow (20~km~s$^{-1}$) non-dissociative shock and a 
faster (30--60~km~s$^{-1}$) dissociative shock are required 
to explain our ISO spectra. The density of the pre-shock gas 
in Cep~A East is similar to that in the S106 PDR: 
10$^5$--10$^6$~cm$^{-3}$. Our J-shock component may be 
identified with the [Fe\,{\sc ii}] 1.64~$\mu$m clumps 
observed by Goetz et al. (1998). The C-shocks could be 
due to their H$_2$ knots which do not show ionic emission.

It is interesting to compare the results derived here for the emission 
from the eastern lobe of Cep~A with the results by Wright et al. (1996), 
obtained using the same instruments on board ISO, for the western part of 
the nebula. They derived a $T_{\rm rot}$ of 700 $\pm$ 30~K for the H$_2$ 
lines with upper level 
energies up to 7000~K, and a temperature range up to 11,000~K for the higher 
upper level energies. They explained this H$_2$ emission as arising from a 
combination of at least two C-shocks with different pre-shock density, shock 
velocity and covering factor. In addition to this, they also reported 
emission from [Ne\,{\sc ii}], [S\,{\sc i}] and [Si\,{\sc ii}], which they 
explained as arising in a planar J-shock with pre-shock density of 
10$^3$--10$^4$~cm$^{-3}$ and shock velocity 70--80~km~s$^{-1}$. 
Qualitatively, the detected lines in the eastern and western part of the 
nebula are well in agreement. The main difference 
between the observed lines seems to be that they are much more intense in 
the western part of Cep~A, indicating that the densities there are a 
factor of 100 higher than those obtained for the eastern lobe, but the 
shock velocities are comparable in both parts of the nebula. The similarity 
between these two parts of Cep~A are in agreement with a scenario in which 
the driving source of the molecular outflow, HW-2, went through multiple 
episodes of outflow activity, as suggested by Narayanan \& Walker (1996). 
In this scenario, the J-shock component could be due to the most recent 
period of enhanced mass loss, whereas one or more C-shocks could be due 
to older generations of outflows. Alternatively, the stellar wind material 
could produce a (fast) J-shock while the surrounding molecular 
cloud might be swept up by a slower C-shock.

As was already remarked by Staude \& Els\"asser (1993), the differences 
between the environments of these two massive embedded YSOs, 
S106 and Cep~A, are remarkable. The mid- and far-infrared observations 
presented in this paper show that the difference between these two 
sources cannot be due to a different orientation of their nebulae; 
it can only be a reflection of their different evolutionary status, 
with Cep~A being the younger of the two. 
In S106 the stellar wind and UV radiation 
of the exciting source have cleared and excited a sufficiently large region 
to create strong PDR emission, whereas in the case of Cep~A, the central 
source is still heavily embedded and we only observe the interaction of 
the outflow with its surroundings. In due time, Cep~A will clear its 
surroundings, ionize the hydrogen, and evolve into a bipolar nebula 
quite similar to S106.

\acknowledgements{The authors would like to thank John Rayner and Klaus-Werner 
Hodapp for providing us with the K-band images of S106 and Cep~A shown in 
Figs.~1 and 2. Peter van Hoof is kindly acknowledged for providing us with a copy 
of the {\sc cloudy} computer code. Frank Molster is thanked for providing the 
authors with valuable input on a draft of the paper. MvdA acknowledges financial 
support from NWO grant 614.41.003 and through a NWO {\em Pionier} grant to 
L.B.F.M. Waters. This research has made use of the Simbad data base, operated 
at CDS, Strasbourg, France.}


\begin{thebibliography}{}
\bibitem{}
Alonso-Costa, J.L., Kwan, J. 1989, ApJ 338, 403
\bibitem{}
Bally, J., Lane, A.J. 1982, ApJ 257, 612
\bibitem{}
Bally, J., Yu, K.C., Rayner, J., Zinnecker, H. 1998, AJ 116, 1868
\bibitem{}
Bieging, J.H. 1984, ApJ 286, 591
\bibitem{}
Black, F.H., van Dishoeck. E.F. 1987, ApJ 322, 412
\bibitem{}
Boogert, A.C.A., Schutte, W.A., Tielens, A.G.G.M., et al.,
 1996, A\&A 315, L377
\bibitem{}
Boogert, A.C.A., Schutte, W.A., Helmich, F.P., et al.,  
 1997, A\&A 317, 929
\bibitem{}
Burton, M.G., Hollenbach, D.J., Tielens, A.G.G.M. 1990, ApJ 365, 620
\bibitem{}
Burton, M.G., Hollenbach, D.J., Tielens, A.G.G.M. 1992, ApJ 399, 563
\bibitem{}
Casement, L.S., McLean, I.S. 1996, ApJ 462, 797
\bibitem{}
Cesarsky, D., Lequeux, J., Abergel, A., Perault, M., Palazzi, E., 
 Madden, S., Tran, D. 1996, A\&A 315, L309
\bibitem{}
Chernoff, D.F., Hollenbach, D.J., McKee, C.F. 1982, ApJ 259, L97
\bibitem{}
Clegg, P.E., Ade, P.A.R., Armand, C., et al., 1996, A\&A 315, L38
\bibitem{}
Dabrowski, I. 1984, Canadian J. Phys. 62, 1639
\bibitem{}
Day, K.L. 1979, ApJ 234, 158
\bibitem{}
de Graauw, Th., Haser, L.N., Beintema, D.A., et al., 1996, 
 A\&A 315, L49
\bibitem{}
d'Hendecourt, L., Jourdain de Muizon, M., Dartois, A., Breitfellner, M., 
 Ehrenfreund, P., Benit, J., Boulanger, F., Puget, J.L., Habing, H.J. 1996, 
 A\&A 315, L365
\bibitem{}
Doyon, R., Nadeau, D. 1988, ApJ 334, 883
\bibitem{}
Draine, B.T., Bertoldi, F. 1996, ApJ 468, 269
\bibitem{}
Drew, J.E., Bunn, J.C., Hoare, M.G. 1993, MNRAS 265, 12
\bibitem{}
Eiroa, C., Els\"asser, H., Lahulla, J.F. 1979, A\&A 74, 89
\bibitem{}
Ellis, H.B., Lester, D.F., Harvey, P.M., Joy, M., Telesco, C.M., Decher, R., 
 Werner, M.W. 1990, ApJ 365, 287
\bibitem{}
Evans, N.J., Becklin, E.E., Beichman, C., Gatley, I., Hildebrand, R.H., 
 Keene, J., Slovak, M.H., Werner, M.W., Whitcomb, S.E. 1981, ApJ 244, 115
\bibitem{}
Felli, M., Staude, H.J., Reddmann, T., Massi, M., Eiroa, C., Hefele, H., 
 Neckel, T., Panagia, N. 1984, A\&A 135, 261
\bibitem{}
Felli, M., Simon, M., Fischer, J., Hamann, F. 1985, A\&A 145, 305
\bibitem{}
Ferland, G.J. 1996, Univ. of Kentucky Physics Department Internal Report
\bibitem{}
Fluks, M.A., Plez, B., Th\'e, P.S., de Winter, D., Westerlund, B.E., 
 Steenman, H.C. 1994, A\&AS 105, 311
\bibitem{}
Garden, R.P., Geballe, T.R. 1986, MNRAS 220, 611
\bibitem{}
Geballe, T.R., Lacy, J.H., Persson, S.E., McGregor, P.J., Soifer, B.T. 
 1985, ApJ 292, 500
\bibitem{}
Gehrz, R.D., Grasdalen, G.L., Castelaz, M., Gullixson, C., Mozurkewich, D., 
 Hackwell, J.A. 1982, ApJ 254, 550
\bibitem{}
Gerakines, P.A., Schutte, W.A., Greenberg, J.M., van Dishoeck, E.F. 
 1995, A\&A 296, 810
\bibitem{}
Goetz, J.A., Pipher, J.L., Forrest, W.J., Watson, D.M., Raines, S.N., 
 Woodward, C.E., Greenhouse, M.A., Smith, H.A., Hughes, V.A., Fischer, J. 
 1998, ApJ 504, 359
\bibitem{}
Greene, T.P., Rayner, J.T. 1994, in ``Astronomy with Infrared Arrays'', 
 ed. I. McLean, Kluwer Academic Publishers, p. 453
\bibitem{}
Habing, H.J. 1968, Bull. Astron. Inst. Netherlands 19, 421
\bibitem{}
Hartigan, P., Carpenter, J.M., Dougados, C., Skrutskie, M.F. 1996, AJ 111, 1278
\bibitem{}
Hayashi, S.S., Hasegawa, T., Tanaka, M., Hayashi, M., Aspin, C., McLean, I.S., 
 Brand, P.W.J.L., Gatley, I. 1990, ApJ 354, 242
\bibitem{}
Hefele, H., H\"olzle, E. 1980, A\&A 88, 145
\bibitem{}
Hippelein, H., M\"unch, G. 1981, A\&A 99, 248
\bibitem{}
Hodapp, K.W. 1994, ApJS 94, 615
\bibitem{}
Hodapp, K.W., Rayner, J. 1991, AJ 102, 1108
\bibitem{}
Hollenbach, D.J., McKee, C.F. 1989, ApJ 342, 306
\bibitem{}
Hollenbach, D.J., Tielens, A.G.G.M. 1999, Rev. Mod Phys. 71, 173
\bibitem{}
Hughes, V.A. 1988, ApJ 383, 280
\bibitem{}
Hughes, V.A., Wouterloot, J.G.A. 1984, ApJ 276, 204
\bibitem{}
Kaufman, M.J., Neufeld, D.A. 1996, ApJ 456, 611
\bibitem{}
Kessler, M.F., Steinz, J.A., Anderegg, M.E., et al., 1996, A\&A 315, L27
\bibitem{}
Kirby-Docken, K., Liu, B. 1978, ApJS 36, 359
\bibitem{}
Kurucz, R.L. 1991, in ``Stellar atmospheres--Beyond classical models''
 (eds. A.G. Davis Philip, A.R. Upgren, K.A. Janes), L. Davis press, 
 Schenectady, New York, p. 441
\bibitem{}
Lenzen, R., Hodapp, K.W., Solf, J. 1984, A\&A 137, 202
\bibitem{}
Longmore, A.J., Robson, E.I., Jameson, R.F. 1986, MNRAS 221, 589
\bibitem{}
Loushin, R., Crutcher, R.M., Bieging, J.H. 1990, ApJ 362, L67
\bibitem{}
Maillard, J.P., Joblin, C., Mitchell, G.F., Vauglin, I., Cox, P. 
 1999, in ``The universe as seen by ISO'', eds. P. Cox \& M.F. Kessler, 
 ESA SP-427, p. 707.
\bibitem{}
Mel'nikov, S.Y., Shevchenko, V.S., Grankin, K.N. 1995, Astron. Rep. 39, 42
\bibitem{}
Mezger, P.G., Chini, R., Kreysa, E., Wink, J. 1987, A\&A 182, 127
\bibitem{}
Narayanan, G., Walker, C.F. 1996, ApJ 466, 844
\bibitem{}
Parmar, P.S., Lacy, J.H., Achtermann, J.M. 1991, ApJ 372, L25
\bibitem{}
Persson, S.E., McGregor, P.J., Campbell, B. 1988, ApJ 326, 339
\bibitem{}
Pipher, J.L., Sharpless, S., Savedoff, M.P., Kerridge, S.K., Krassner, J., 
 Schurmann, S., Soifer, B.T., Merrill, K.M. 1976, A\&A 51, 255
\bibitem{}
Rayner, J. 1994, in ``Infrared Astronomy with Arrays: The next generation'', 
 ed. I.S. McLean (Dordrecht: Kluwer), p. 185
\bibitem{}
Schneider, N., Kramer, C., Simon, R. 2000, in Proc. 33$^{\rm rd}$ ESLAB Symposium 
 ``Star Formation from the Small to the Large Scale'', eds. F. Favata, A.A. Kaas 
 \& A. Wilson, ESA SP-445, in press
\bibitem{}
Schutte, W.A., Tielens, A.G.G.M., Whittet, D.C.B., Boogert, A.C.A., 
 Ehrenfreund, P., de Graauw, Th., Prusti, T., van Dishoeck, E.F., Wesselius, 
 P.R. 1996, A\&A 315, L333
\bibitem{}
Staude, H.J., Els\"asser, H. 1993, Astron. Astroph. Rev. 5, 165
\bibitem{}
Staude, H.J., Lenzen, R., Dyck, H.M., Schmidt, G.D. 1982, ApJ 255, 95
\bibitem{}
Tanaka, M., Hasegawa, T., Hayashi, S.S., Brand, P.W.J.L., Gatley, I. 1989, 
 ApJ 336, 207
\bibitem{}
Tielens, A.G.G.M., Allamandola, L.J. 1987, in ``Interstellar Processes'', 
 eds. D.J. Hollenbach \& H.A. Thronson Jr. (Dordrecht: Reidel), p. 397
\bibitem{}
Tielens, A.G.G.M., Hollenbach, D.J. 1985, ApJ 291, 722
\bibitem{}
Turner, J., Kirby-Docken, K., Dalgarno, A. 1977, ApJS 35, 281
\bibitem{}
Whittet, D.C.B., Schutte, W.A., Tielens, A.G.G.M., et al., 
 1996, A\&A 315, L357
\bibitem{}
Wright, C.M., Drapatz, S., Timmermann, R., van der Werf, P.P., Katterloher, 
 R., de Graauw, Th. 1996, A\&A 315, L301
\end{thebibliography}
\end{document}